\documentclass{aa}  
\usepackage{graphicx}
\usepackage{txfonts}

\usepackage{color, colortbl}
\usepackage[hidelinks]{hyperref}
\usepackage{xcolor}
\hypersetup{
    colorlinks = true, 
    linkcolor={red!50!black},
    citecolor={blue!50!black},
    urlcolor={blue!80!black}
}
\usepackage{amstext}

\begin{document} 

\definecolor{bubbles}{rgb}{0.91, 1.0, 1.0}
\definecolor{columbiablue}{rgb}{0.61, 0.87, 1.0}
\definecolor{cream}{rgb}{1.0, 0.99, 0.82}
\definecolor{lightblue}{rgb}{0.68, 0.85, 0.9}
\definecolor{lightcyan}{rgb}{0.88, 1.0, 1.0}

   \title{Digging Deeper for RR Lyrae Stars with Low Modulation Amplitudes}
   \author{Geza Kovacs 
          }

   \institute{Konkoly Observatory, Research Center for Astronomy and Earth Sciences  
              of HUN-REN \\ 
              Budapest, 1121 Konkoly Thege ut. 15-17, Hungary\\
              \email{kovacs@konkoly.hu}
             }

   \date{Received 22-04-2025 / Accepted DD-MM-2025}

%
%
%
  \abstract
{With the goal of searching for very low modulation amplitudes among 
fundamental mode RR~Lyrae stars and assess their incidence rate, we 
performed a survey of $36$ stars observed by the Kepler satellite 
during the entire four-year period of its mission. The search was 
conducted by a task-oriented code, designed to find low-amplitude 
signals in the presence of high-amplitude components and instrumental 
systematics.
We found $7$ new modulated stars and negate one earlier 
claimed star, whereby increasing the number of known Blazhko stars 
from $17$ to $24$ and yielding an observed occurrence rate of $67$\% 
for the Kepler field. 
Six of the new stars have the lowest modulation amplitudes found so 
far, with $\sim 250$~ppm Fourier side-lobe amplitudes near the 
fundamental mode frequency. 
Because of the small sample size in the Kepler field, 
we extended the survey to $12$ campaign fields observed by K2, the 
``two-wheeled'' mission of Kepler. From the $1061$ stars we found 
$514$ Blazhko stars. After correcting for the short duration of the 
time spent on each field, and for the noise dependence of the detections, 
we arrived at an underlying occurrence rate of $\sim 75$\% -- likely a 
lower limit for the true rate of Blazhko stars in the K2 fields.}


\titlerunning{Digging deeper for Blazhko stars}
\authorrunning{Kovacs}

   \maketitle
%
%
%
\section{Introduction}
\label{sect:intro}
RR~Lyrae stars are unique objects among the large-amplitude variables.  
At least half of the fundamental mode (RRab) stars show periodic light 
curve modulation, commonly known as Blazhko phenomenon -- see 
\cite{blazhko1907}, \cite{shapley1916} and \cite{szeidl2000} for some 
historical background. The modulation amplitudes vary in a wide range 
from object to object, occasionally reaching the size of the amplitude 
of the pulsation. Large number of observational works have been devoted 
to these stars during the past hundred years, without leading to any 
clue to the root cause of this mesmerizing physical phenomenon.    

Among the various observables, that could play role in the discovery 
of the basic physics behind the modulation, here we embark on the 
frequency of the Blazhko objects. Because measurement accuracy and 
data volume (especially length of the timebase) obviously play an 
important role in the detectability of the modulation, we decided to 
use the single field data from the nearly four-year long Kepler mission, 
and that of the extended mission 
K2\footnote{\url{https://science.nasa.gov/mission/kepler/in-depth/}},  
covering various fields but with a much shorter duration of $\sim 80$ 
days each. Although both of these sets are of (very) high quality, 
the high stellar density in the Kepler field is a serious issue, in 
particular, when the various quarters are to be merged for a joint 
analysis. Similarly, in the light of the analysis of the OGLE Galactic 
Bulge data by \cite{prudil2017} and \cite{skarka2020}, the K2 data 
surely miss many objects with long-period modulations. In spite of 
these, and some other drawbacks, such as the low sample size for the 
Kepler data and the overwhelming instrumental noise for many stars 
in the K2 data, these sets obviously belong to the non-missable data 
domain in the present context.  

Observed occurrence rates reported so far for modulated RRab stars from  
various stellar populations range from the ridiculously low $12\%$ 
\citep[field stars in the Large Magellanic Cloud, via the MACHO survey 
-- see ][]{alcock2003} to the shockingly high $\sim 90\%$ 
\citep[Galactic field, from the analysis of several K2 fields by][]{kovacs2018,kovacs2021}. 
The importance of the sample size and, in particular, of the data 
precision, can be further illustrated by the change from the earlier 
estimate of $20-30\%$ for Galactic field stars \citep[][]{szeidl1988}  
to the almost $50\%$, estimated later by \cite{jurcsik2009} from a small 
(i.e., $30$) but high-quality sample. Similarly, for the Galactic Bulge 
from a relatively small sample of $150$ stars from the OGLE-I inventory, 
\cite{moskalik2003} derive $\sim 20\%$, whereas from the accumulated data 
of $\sim 8000$ stars by the domination of the observations made in the 
OGLE-IV phase, \cite{prudil2017} arrives to  $\sim 40\%$. 

Here, focusing solely on RRab stars and following the methodology 
to be described in Sect.~\ref{sect:method}, we perform a frequency 
analysis for the combined data of nearly all quarters of the original 
Kepler mission. We also extend our earlier work on the K2 campaign 
fields and search for RR~Lyrae stars with low modulation amplitudes. 
The basic question we intend to address is this: Is there a real 
dearth of low-amplitude Blazhko stars or the data are still insufficient 
to address this question? In other words: are all non-detections 
attributable to genuine non-modulated monoperiodic fundamental-mode 
pulsators? 

%
%
\section{Method}
\label{sect:method}
\begin{figure}[h]
\centering
\includegraphics[width=0.48\textwidth]{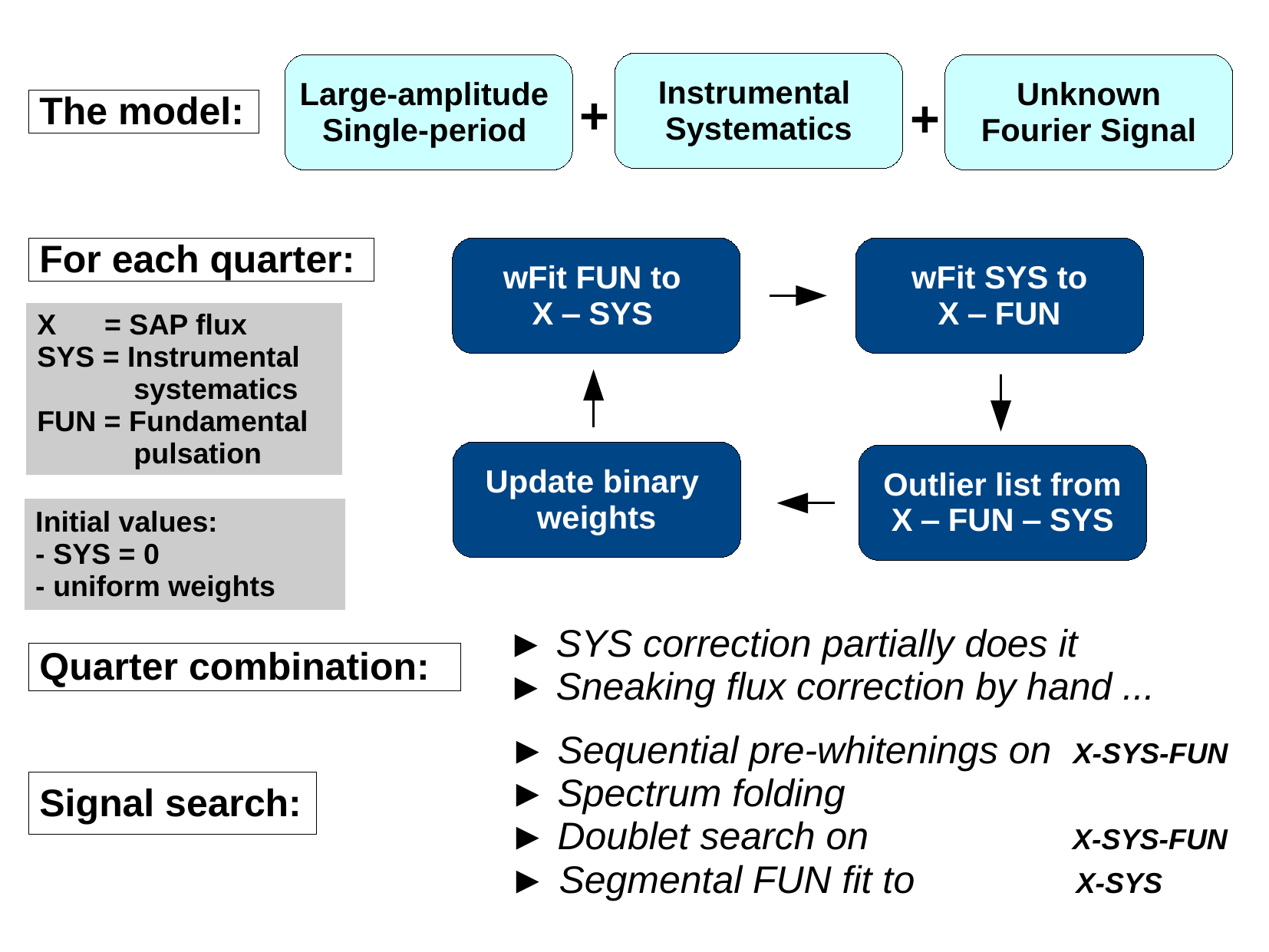}
\caption{Brief summary of the time series procedures employed 
in this work to search for modulated RR~Lyrae stars. The prefix 
``w'' in wFit denotes weighted Least Squares fit, controlled 
by the outlier array, generated during the fitting cycle. Additional 
details can be found in Sect.~\ref{sect:method}} 
\label{flow_chart}
\end{figure}

With some important modifications, here we essentially follow the 
methodology used in our former papers dealing with modulated RR~Lyrae 
stars \citep{kovacs2018, kovacs2021}. The basic steps and signal 
constituents are depicted in Fig.~\ref{flow_chart}. In the following we 
describe the most important features of the data handling and 
analysis.\footnote{The actual physical implementation is realized by a 
stand alone Fortran code, relying only on the publicly available raw 
photometric time series.}   

%
%

{\em Signal Component Separation:} 
The fundamental problem in discovering a shallow signal in the 
presence of a large signal ({\sf FUN}) and instrumental systematics 
({\sf SYS}) is the separation of these constituents with a minimum 
distortion of the components, including the unknown component. The 
separation of the components is executed in a simple iteration loop. 
Because the shallow signal is unknown, we cannot include it in the 
iteration, except if, at each step of the iteration, a frequency 
analysis is performed on the residuals {\sf X-FUN-SYS}, where 
{\sf X} denotes the input signal, i.e., the raw data, usually obtained 
from simple aperture photometry (SAP). This, 
however, could be extremely expensive computationally, and, based 
on our earlier efforts, does not seem to be profitable in terms of the 
discovery of new signals. Therefore, only {\sf FUN} and {\sf SYS} 
are involved in the iteration, and then, when the iteration is 
completed, the residuals {\sf X-FUN-SYS} are searched for new signals. 

{\em The Systematics Correction Vector:} 
We employ co-trending to filter out instrumental systematics via 
simple weighted least-squares. Co-trending is based on the assumption 
that systematics can be modeled by using the linear combination of 
non-variable stars measured simultaneously in the same field where 
the target is \citep[][]{kovacs2005, tamuz2005, smith2012}. The 
correction time series (or vectors) are built up from a properly 
selected set of non-variable stars. Then, this set is run through 
some standard dimension reduction method 
\citep[e.g., Singular Value Decomposition, or SVD -- see][]{press1992}, 
whereby a new set of vectors 
is constructed from all the vectors selected in the first step. 
These SVD vectors are ordered according to ``essentiality'' and 
some fraction of them is used as {\sf SYS}. For the Kepler data we 
employ some $10$--$30$, whereas for K2 we use $100$ SVD vectors 
(constructed from some $\sim 500$ stars in each field). 
It is important to note that target-dependent external parameters 
\citep[such as photocenter pixel positions -- see][]{bakos2010} 
are not used in this work, because they may lead to considerable 
signal distortion and do not result in a higher signal detection rate.
 
{\em Quarterly fit of {\sf FUN} and {\sf SYS}:} 
For the systematics correction vectors {\sf SYS} it is clear 
that the fit is performed on a quarterly basis. However, 
when global functions (functions, that are assumed to be independent 
of the quarters) are also included in the model, this is no 
longer true. In general, the signals, we are searching for, 
are in the ppt -- sub-ppt regime. Our experimentation with the 
model  including {\sf FUN} as a global function, suggests that 
this approach always leaves extra power very close to the fundamental 
mode frequency $f_0$ and its harmonics, making the detection 
of the nearby faint signal components more difficult. Consequently, 
we opted to the quarterly-based {\sf FUN} fit, and assume that 
the very small differences among the so-obtained solutions for 
{\sf FUN} reflect the remaining small systematics, not filtered 
out by the previous systematics correction steps. Obviously, if 
the modulation has a small amplitude and long period, this method 
will suppress the modulation and this particular property of the 
target will remain hidden. Nevertheless, the quarterly separate fits 
are in line with the way how we treat the K2 time series, and 
short-period Blazhko phenomena (with $P_{\rm BL}\lse 50~days$) 
can still be very well studied and get warning signs at moderately 
longer periods. 
 
{\em Outlier Selection:} 
In many cases outliers could cause troubles if they are not 
treated correctly. After some experimentation with a robust fit 
algorithm, we opted for a ``milder'' approach by using a simple 
sigma-clipping method during the iterative phase of the systematics 
filtering. It is important to note that we select the outliers from 
{\sf X-FUN-SYS} and start with uniform weights. Once some outlier 
label array is generated, we use these in the weighted least squares 
fits. With a properly chosen sigma-clipping factor (usually between 
$3.5$ and $5$ standard deviations), at each iteration the weight 
array is updated with $0$ or $1$, depending on the current outlier 
status of the given data item.    

{\em Quarter Stitching:} 
For the Kepler data, where seasonal segments (quarters) are available, 
a special issue is the combination of those segments. Because the 
leading cause of the different flux variations from quarter to 
quarter is the sneaking in and out neighboring stars in the aperture 
where the raw flux is measured, instead of using two proper flux-adjusting 
parameters \citep[e.g.,][]{nemec2011,benko2014}, we opted for a single 
parameter that is constant throughout the full span of a given quarter 
but varies between the different quarters, and, of course, it is 
also target dependent. The sneaking flux correction enters in the 
raw (i.e., {\sf SAP}) flux as follows 
%
%
\begin{eqnarray}
\label{sneak_corr}
X_{corr}(i) = 2.5\log(X_{raw}(i)+F_{sneak}), \hspace{2mm}
\end{eqnarray}
where $X_{raw}(i)$ is the $i$-th input (actually measured) stellar 
flux and $F_{sneak}$ is the properly adjusted sneak flux for a 
given target and a given quarter. Because the systematics corrections 
quarter-dependent, most of the adjustment are already done via the 
iterated {\sf SYS} time series. $F_{sneak}$ is needed to tackle 
the uncorrected part. The value of $F_{sneak}$ is selected on 
a trial and error basis after inspecting either the multi-quarter, 
systematics-corrected signal {\sf X-SYS}, or the variation of the 
temporal Fourier fits (see below). 

{\em Modulation Diagnostics:} 
The basic diagnostic tool is the 
\underline{prewhitened Fourier spectrum}\footnote{If not stated 
otherwise, the word ``prewhitened'' is used for the data/spectra 
obtained after subtracting the fundamental mode {\sf FUN} from the 
systematics-filtered data {\sf X-SYS}.} 
\citep[DFT -- Discrete Fourier Transform,][]{deeming1975}. The lack 
of residual power near the fundamental mode frequency $f_0$ and/or 
its harmonics disqualifies any further inquiry about a possible 
modulation component.\footnote{This is slightly an ``over safe'' 
condition, because the doublet search method (see later) is, in 
general, more sensitive to modulation components due to power 
accumulation from several harmonics. DFT yields information only 
separately on these components, and therefore, the signal-to-noise 
ratio is smaller.} 
Assuming repeating frequency patterns around $f_0$ and its harmonics, 
the above spectrum can be used to compute the 
\underline{folded Fourier spectrum}. Here we simply add up the 
power (i.e., the square of the amplitudes obtained from DFT)   
in the $\pm f_0/2$ neighborhood of $f_0$ and its harmonics throughout 
the evaluated spectrum. We use the square root of the so-derived 
power spectrum to maintain compatibility with the original DFT 
spectrum. In the method of 
\underline{Doublet search}, we assume that the prewhitened time 
series {\sf X-FUN-SYS} can be approximated by frequency doublets 
around $f_0$ and some number of harmonics (here we use frequencies 
up to the $5^{th}$ harmonics). By changing the modulation frequency 
$f_{mod}$, we fit these doublets to {\sf X-FUN-SYS}, and compute 
the RMS of the fit and search for the minimum value. Although the 
exactly repeating frequency distribution does not seem to be a rule 
in all cases, this method has been proven the most powerful diagnostics 
to detect very shallow modulation components, barely seen in the 
DFT spectrum. Finally, following \cite{nemec2011}, in the method of 
\underline{Segmental {\sf FUN} fit}, we use overlapping small 
segments of the full light curve {\sf X-SYS} (corrected for sneaking 
fluxes in multi-quarter case) to fit {\sf FUN} and use the temporal 
Fourier amplitudes and phases ($A_1(t)$ and $\varphi_1(t)$, respectively) 
at $f_0$. The pattern observed in these Fourier parameters is also 
a very powerful diagnostic tool and weights in the final evaluation of 
a given target. 

\begin{figure*}[t]
\centering
\includegraphics[width=0.80\textwidth]{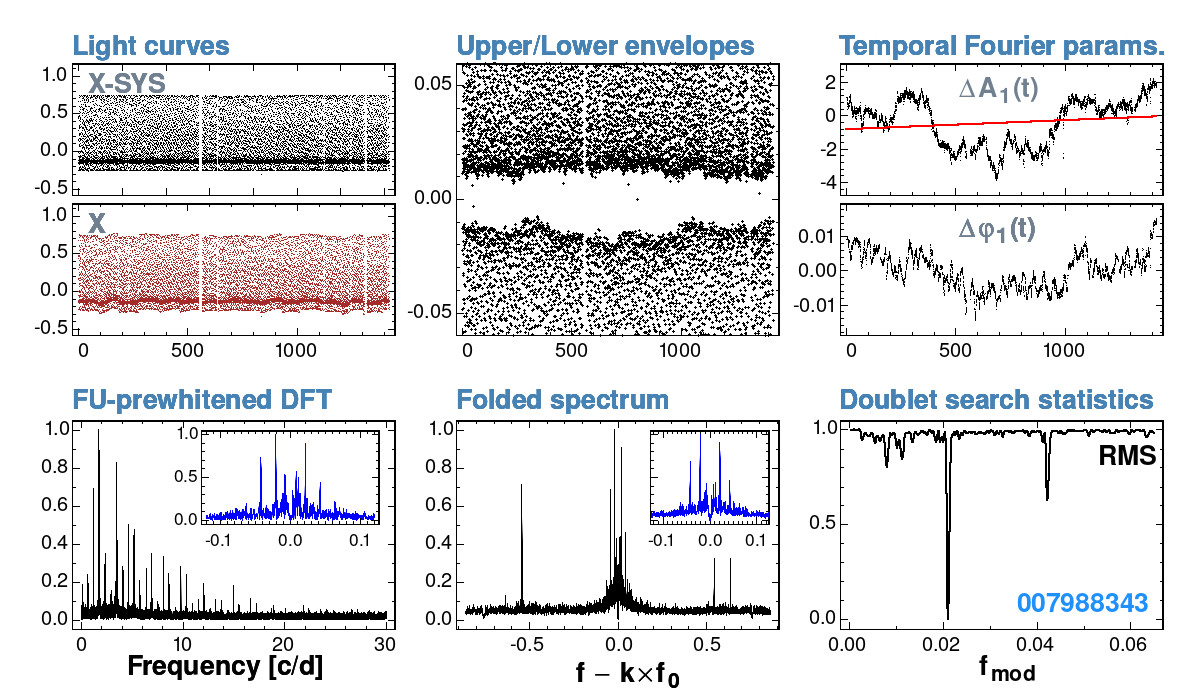}
\caption{An example of the diagnostic diagrams for an RR Lyrae 
         star from the Kepler field used in search for modulated 
	 stars. 
	 {\em Upper row, from left to right:}
	 systematics corrected (upper sub-panel) and raw light 
	 curves, units are days and relative magnitudes; 
	 close-up of the lower and upper parts (respectively, 
	 upper and lower halves of the panel) of the filtered 
	 light curve shown in the first panel (units are the same 
	 as in the first panel); variation (relative to the average) 
	 of the time-dependent Fourier parameters at the fundamental 
	 frequency $f_0$. Red line: linear regression to $\Delta A_1(t)$. 
	 Vertical axis units: [ppt] and [rad];
	 {\em Lower row, from left to right:}
	 full DFT of the prewhitened data with the inset zooming 
	 on the region of $f_0$ (horizontal axis shows $f-f_0$); 
	 folded spectrum, with similar plot structure as for the 
	 previous panel; 
	 RMS values of the doublet search at various modulation 
	 frequencies. The KIC ID (also known as V1510~Cyg) is shown 
	 in the lower right corner. 
	 All vertical axes are normalized to the peak values, horizontal 
	 axes have the same units (i.e., [c/d]). See Sect.~\ref{sect:method} 
	 for additional details.} 
\label{Kepler_diagnostics}
\end{figure*}

{\em Simple Light Curve inspection:}
Serving as an overall data quality assessment, all the above can be 
combined by inspecting the full light curve (including the raw data) 
and the close-ups of the upper/lower envelopes of the systematics-filtered 
light curves. Examples of the diagnostic diagrams are shown in 
Figs.~\ref{Kepler_diagnostics}, \ref{K2_diagnostics} in later sections.  

{\em Annual variation test:}
Because the systematics correction vector {\sf SYS} is composed  
from bright stars, they are supposedly less affected by seasonal 
and annual variations than the RR Lyrae stars, that are on the 
fainter tail of the brightness distribution. As a result, in 
general, co-trending is not able to eliminate fully long-term 
systematics exhibiting as fake annual modulations in several 
RR Lyrae stars. To test if suspicious modulations near the time 
scale of a year\footnote{Although the orbital period of the 
Kepler satellite is slightly longer than that of the Earth, 
using the latter does not cause any appreciable difference.} 
and its harmonics may have anything to do with 
some instrumental effect, for the Kepler data we also fit a 
Fourier sum with frequencies 
\{$j\times f_0 \pm k/365 ; j,k=1,2,..,5$ \}, and check if the 
suspected component is eliminated. If so, the candidate is dropped.  

{\em Target Classification:}
We use a simple three level classification scheme for the likelihood 
of a modulation signal\footnote{That is, a close component to the 
fundamental mode, with a frequency difference less than 
$\sim 0.15$ c/d (we note that very few objects fall in the  
frequency interval of $0.10-0.15$ c/d).} being present in a given 
target. The lack of significant residual peaks near $f_0$ and/or its 
harmonics leads to {\bf class 0}. Peaks, close to noise level, may 
also lead to the above classification, if other diagnostics are not 
strong enough for other classification. If the residual peak is strong,  
it usually leads to {\bf class 2}, implying a possible long-period 
modulation, assuming that other diagrams do not suggest otherwise 
(e.g., strongly corrupted data by observational/instrumental 
effects). Finally, if all diagrams are cooperative, and, in 
particular, the doublet search suggests the presence of a 
significant signal, then the object is labeled as {\bf class 1}, 
i.e., an RR Lyrae with Blazhko effect. Additional complications 
may occur during the analysis of the K2 data, when multiple data 
sources are available with different classifications. In this 
case we assign class 1 to the star if at least one of the sources 
has this classification. Otherwise, the star is labeled as  
class 2. This scheme favors decisions made on detections even if 
the data sources used to make these decisions are in minority 
with respect of the remaining sets analyzed for a given target.

%
%
%
%

%
%
\section{Datasets}
\label{sect:data}
We analyzed the RRab stars from two databases: the one gathered by 
the Kepler satellite during its original mission by monitoring a 
single field in a four-year span \citep{borucki2017}; and the 
multi-field data collected by the extended mission K2 \citep{howell2014} 
along the ecliptic belt. While during the original mission the 
pointing was accurate, the K2 phase was overwhelmed by instrumental 
systematics due to correction mechanism applied for the lost 
reaction wheels. In the rest of the paper we refer to the 
data gathered during the original Kepler mission as ``Kepler data''. 
The quarters are referred as Q\#\#, where the hash marks refer to 
the 2-digit quarter numbers. The campaign fields in the case of 
the K2 data is referred similarly, starting with letter ``C''.   

We use only long-cadence (i.e., $\sim$~half-hour-integrated) 
data from both missions. Except for Q01, all available 
quarters are employed. Most of the targets we deal with in the 
case of the Kepler data contain over $60000$ data points, allowing 
(in principle) a search for sinusoidal signals at the level of few 
times $10$~ppm for the RR~Lyrae sample we deal with. Following 
the list of \cite{plachy2021}, a sample of $37$ RRab stars have 
been investigated. Although there is an extension of this list by 
\cite{forro2022}, containing additional, RRab stars ($20$ of them), 
after experimenting with the data, we decided not to use these 
stars. The stars are mostly faint, and therefore, they not only 
have larger noise but they also suffer from stronger systematics 
due to the crowded nature of the Kepler field. With these quality 
factors, it is nearly impossible to discover hidden low-amplitude 
Blazhko stars (in particular those with long modulation periods). 
All data used for the analysis of the Kepler data have been 
downloaded from the STSci MAST 
site.\footnote{\url{https://archive.stsci.edu/kepler/data_search/search.php}}  

The RRab list provided by \cite{molnar2018,plachy2019} were 
used to select the targets for the K2 analysis. Although there 
is an updated, more extensive list by \cite{bodi2022}, we 
decided not to use their list (and the accompanying data) 
in the present analysis, because the set based on the above two 
lists is large enough to ensure statistical stability of the 
results derived from this sample. The campaign fields covered 
by \cite{molnar2018} span from C00 to C13 and includes $1211$  
RRab stars. Extended by the list of \cite{plachy2019}, we start  
with the set of $1301$ stars. Except for C00, all fields are used. 
Finally, we end up with $1061$ stars, considering data availability 
and quality. 

Fortunately, there are several data sources with published raw 
photometry available for the K2 fields. First of all, the 
pipeline of the mission provides an easy access to these data 
via the MAST site. Then, a number of groups made serious efforts 
to treat the data with various methodologies. These works 
resulted in different approaches to derive better raw and 
systematics-filtered photometric data. We use the following 
sources. \cite{petigura2015} -- downloaded from the NASA ExoFop 
site;\footnote {\url{https://exoplanetarchive.ipac.caltech.edu/}}  
\cite{vanderburg2014} and \cite{luger2016,luger2018} -- both have been 
downloaded from the corresponding MAST sites. So, when the K2 
data are analyzed, we have four independent raw time series 
to use as inputs, and therefore, the final classification 
will certainly stand on a more solid basis than if we used 
only a single source. Nevertheless, for various reasons 
(individual project preference, failed target aperture, etc.) 
there are cases when not all four sources are available.

%
%
\section{Results: the Kepler data}
\label{sect:kepler}
Following the method described in Sect.~\ref{sect:method}, we analyzed 
all $37$ RRab stars as listed by \cite{plachy2021}. We quickly detected 
the $18$ previously claimed Blazhko stars but after many repeated 
analyses of the full sample, it turned out that KIC 007021124 did not 
pass our detection criteria, in particular, the quarter stitching had 
led to seasonal changes and no other significant variations were 
identifiable in any of the diagnostic diagrams. Earlier classification 
of this object as a Blazhko star by \cite{benko2015} is based 
predominantly on apparent long-term phase and light curve shape 
variations, which, due to its long-term nature, is hard to disentangle 
from the seasonal variations present in many RR Lyrae stars in 
the Kepler data. Therefore, we labeled this star as class 2.  

Turning to the $19$ stars classified so far as non-Blazhko stars, we 
found that, except for KIC 006100702, all show residual power near 
$f_0$ and its harmonics. This object is bright ($Kp=13.46$), seemingly 
not affected by long-term systematics and lacks any other secondary 
signal component. Therefore, this looks like a genuine single-mode RRab 
star.  
 
Another star, KIC 003866709 turned out to be intractable by the method 
we use. There are repeating flip-flopping systematics from one quarter 
to the other that are target specific, and apparently do not share 
enough commonality with the co-trending set we apply. As a result, no 
classification is possible for this target within our scheme. 
Consequently, we excluded this target from the sample. 

The remaining $18$ stars went through a deep scrutiny by checking the 
various features as discussed in Sect.~\ref{sect:method} and shown 
in Fig.~\ref{Kepler_diagnostics}, exhibiting one of our strongest 
detections. Because the temporal Fourier analysis works on the 
full (systematics-corrected and stitched) light curve, it is 
sensitive to local amplitude and phase variations (see upper right 
panel). The frequency spectra (and also the doublet search -- see 
lower panels), however, work on the quarter-by-quarter 
{\sf FUN}-prewhitened data, and therefore, naturally, they are less 
sensitive to long-term amplitude variations (this means also, that 
they miss such a variation more easily). The low-amplitude 
modulation is best detected by the doublet-fit, but standard DFT 
spectra are also exhibit high-power signals 
\citep[including the infamous ``$f_{68}$'' component -- see, e.g., ][]{benko2019}.

%
\begin{table*}[t]
\begin{flushleft}
\caption{New Blazhko stars from the Kepler field}
\label{tab:new_kep_BL}
\scalebox{0.88}{
\begin{tabular}{clcccccccccc}
\hline
   KIC    & Other     &    Kp  &   N   &    RMS   &  $f_0$   &$A_{tot}$ &  $A_1$   &$f_{mod}$ &$A_{mod}$ & $A_{-}$  & $A_{+}$ \\ 
\hline\hline
003733346 & NR Lyr    & 12.684 & 63808 & 0.001381 & 1.466218 & 0.751480 & 0.260732 & 0.013254 & 0.003640 & 0.000327 & 0.000026\\
006936115 & FN Lyr    & 12.876 & 63798 & 0.001920 & 1.896099 & 1.049814 & 0.373965 & 0.009915 & 0.002365 & 0.000098 & 0.000256\\
009591503 & V894 Lyr  & 13.293 & 63799 & 0.003414 & 1.750129 & 1.105700 & 0.386167 & 0.018225 & 0.003746 & 0.000170 & 0.000115\\ 
009947026 & V2470 Cyg & 13.300 & 63789 & 0.001210 & 1.822856 & 0.602174 & 0.218600 & 0.049724 & 0.001123 & 0.000039 & 0.000078\\ 
007988343 & V1510 Cyg & 14.494 & 63814 & 0.002735 & 1.720748 & 0.996977 & 0.343177 & 0.021107 & 0.003084 & 0.000247 & 0.000234\\
009658012 & --        & 16.001 & 30711 & 0.002793 & 1.875488 & 0.915175 & 0.307399 & 0.049371 & 0.002294 & 0.000297 & 0.000244\\
008344381 & V346 Lyr  & 16.421 & 63796 & 0.008851 & 1.733622 & 0.933623 & 0.320765 & 0.026151 & 0.010011 & 0.000886 & 0.001069\\ 
\hline
\end{tabular}}
\vskip 4 pt
{\bf Notes:}{\small\ 
RMS = standard deviation of the residuals {\sf X-SYS-FUN-BLA}, where {\sf BLA} is 
the frequency doublet approximation of the modulation; $A_{tot}$ = total amplitude 
of {\sf FUN}; $A_1$ = Fourier amplitude at $f_0$; $f_{mod}$ = modulation frequency; 
$A_{mod}$ = total modulation amplitude from {\sf BLA}; $A_{\pm}$ = side-lobe 
Fourier amplitudes at $f_0$. Units: [mag] and [c/d];  
see text on $f_{mod}$ of 009658012 and 009947026.}\\
\end{flushleft}
\end{table*}
%

%
%
\begin{figure}[h]
\centering
\includegraphics[width=0.45\textwidth]{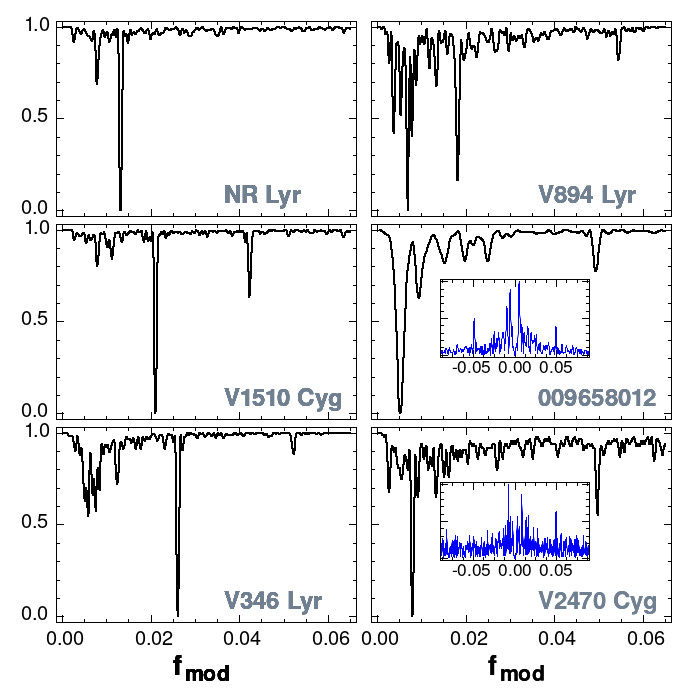}
\caption{Doublet-search statistics for $6$ of the $7$ new Blazhko 
         detections in the Kepler data. Insets show the DFT 
	 spectra of the prewhitened data in the close neighborhood 
	 of the fundamental mode frequency $f_0$ (horizontal 
	 axes are relative to $f_0$). For all panels, the vertical 
	 axes are normalized, frequencies are in [c/d].} 
\label{fig:6new_kep_BL}
\end{figure}

\begin{figure}[h]
\centering
\includegraphics[width=0.45\textwidth]{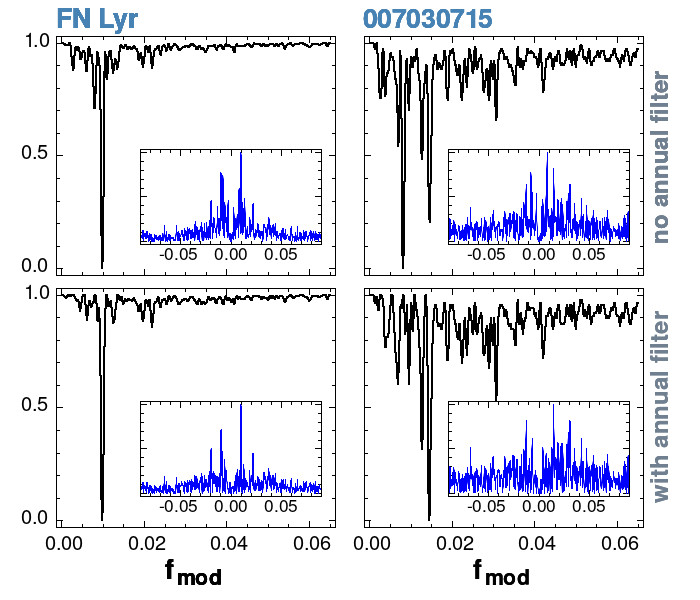}
\caption{{\em Upper row:} as in Fig.~\ref{fig:6new_kep_BL}, 
         for a new class 1 (left) and a class 2 (right) object. 
	 {\em Lower row:} as in the upper row, but the data 
	 were processed by annual filtering to clean the spectrum 
	 from possible instrumental effects on seasonal time scales 
	 (see text and Sect.~\ref{sect:method}).} 
\label{fig:1new_kep_BL}
\end{figure}

Altogether, we identified seven modulated stars among the $18$, 
previously classified as non-Blazhko RRab stars. The doublet search 
diagrams are shown in Figs.~\ref{fig:6new_kep_BL} and \ref{fig:1new_kep_BL}. 
For V894~Lyr we took the clean dip at $f_{mod}=0.018$ c/d, based on 
the annual variation test (see Sect.~\ref{sect:method}) that suppressed 
the low-frequency, barely dominating dip. The same test led to the choice 
of the components near $0.05$~c/d for KIC 009658012 and V2470~Cyg, by 
eliminating the low-frequency component for the latter, or, seriously 
depressing it for KIC 009658012. To compare the doublet search diagrams for 
a class 1 and 2 objects, and illustrate the effect of annual filtering, 
Fig.~\ref{fig:1new_kep_BL} shows the cases of FN~Lyr and KIC 007030715. 
While both spectra have been cleaned by the annual filter, the already 
high signal-to-noise ratio (SNR) detection for FN~Lyr is basically not 
affected, whereas the new dominant dip at $f_{mod}=0.0145$ for 
KIC 007030715 is somewhat tempting. Nevertheless, the SNR of this signal 
is still much too low to consider the object as class 1. 

\begin{figure}[h]
\centering
\includegraphics[width=0.40\textwidth]{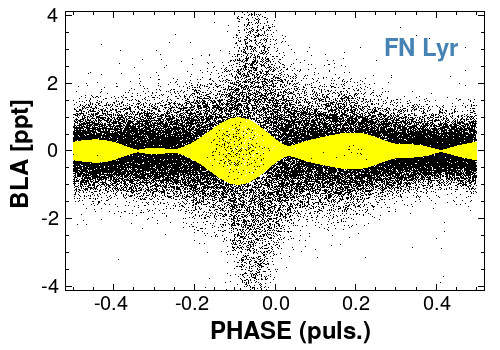}
\caption{Pulsation phase-folded residuals ({\sf X-SYS-FUN}, black dots) 
         for one of the new Blazhko variables in the Kepler data, comprising 
	 nearly $64000$ data points. The maximum light of the pulsation 
	 is at phase zero. The best-fit five-component modulation model 
	 is shown by yellow dots.} 
\label{fig:blaf}
\end{figure}

It may also be a matter of interest to see the actual time series 
corresponding to the modulation component {\sf BLA}. We recall 
that this time series is still a fast oscillating function with 
frequency components $j\times f_0 \pm k\times f_{mod}$. This, together 
with the low amplitude, make it difficult to find a good way to see 
directly the underlying signal. Using temporal Fourier fits and see 
how the low-order Fourier parameters ($A_1$, $\varphi_1$) vary, usually 
works but this plot is affected also by other things, such as the 
combination of the different quarters. Modulations are often displayed 
as residuals on the folded light curve, using the pulsation period as 
the folding period \citep[e.g.,][]{jurcsik2009}. We opt to this method 
and show the corresponding plot for FN~Lyr in Fig.~\ref{fig:blaf}. 
We see that the $5^{th}$-order modulation search model is a good 
approximation of the full {\sf BLA} component, following closely the 
observed pattern of the folded residuals, reaching the highest amplitude 
at the rising branch (just before phase zero) of the large-amplitude 
pulsation. 

%
%
%

%
%
%

%
%
\begin{figure*}[t]
\centering
\includegraphics[width=0.80\textwidth]{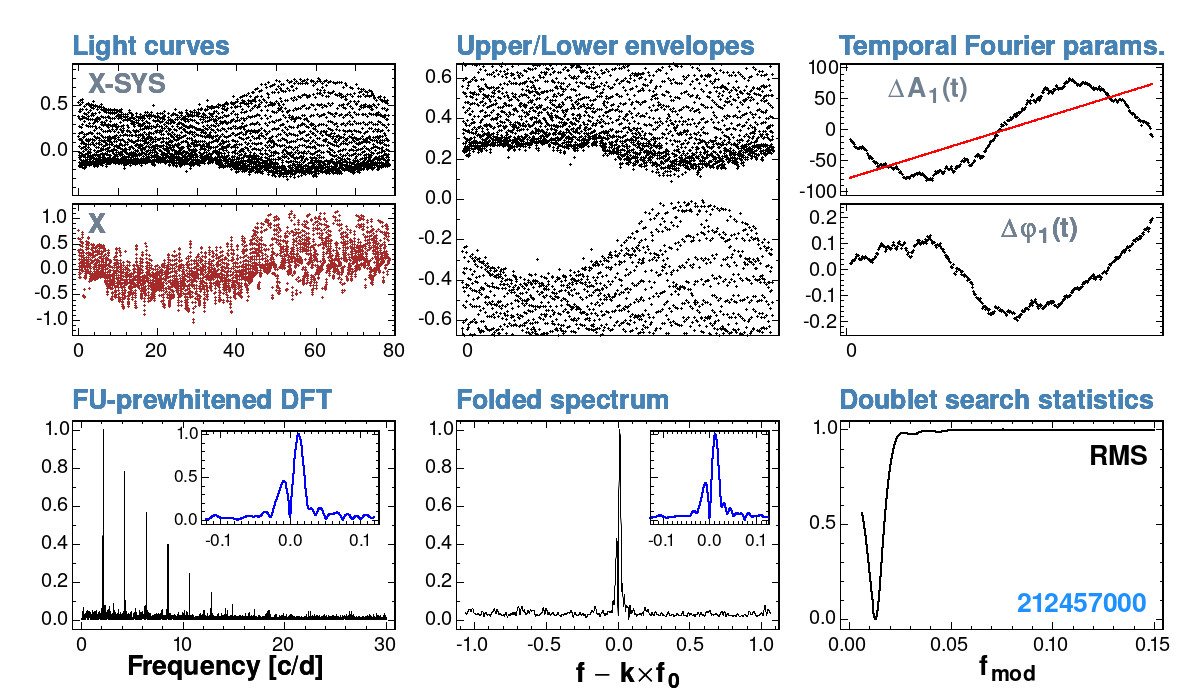}
\caption{This is an example of the diagnostic diagrams obtained from 
         the K2 database. Many stars show strong systematics, such 
	 as EPIC 212457000 from field C06 shown in this figure. 
	 For detailed description of the panels see 
	 Fig.~\ref{Kepler_diagnostics}.} 
\label{K2_diagnostics}
\end{figure*}

The relevant parameters of the seven new Blazhko stars are given in 
Table~\ref{tab:new_kep_BL}. In addition to the explanation given in 
the table notes, we mention the following. The parameters were obtained 
by fitting a Fourier series to {\sf X-SYS}, stitched quarterly. The model 
contains the following frequency triplets: 
\{$j\times f_0-f_{mod}$, $j\times f_0$, $j\times f_0+f_{mod}$ \}, 
with $j$ extending to very high harmonics, usually to 
$\sim 40$.\footnote{Because of the very high data quality, 
leakage to the low frequency regime may happen from components well 
above the Nyquist frequency, from the very high harmonics of the 
fundamental mode frequency $f_0$. This can be avoided by a sufficiently 
high-order Fourier fit of {\sf FUN}.} 
The amplitudes $A_{tot}$ and $A_{mod}$ are peak-to-peak amplitudes 
of the fitted fundamental and Blazhko components ({\sf FUN} and 
{\sf BLA} time series, respectively).

%
%
\section{Results: the K2 data}
\label{sect:k2}
As described in Sect.~\ref{sect:data}, by using four different data 
sources for the raw photometric data, we analyzed campaign fields 
C01-C13 (excluding the microlensing field C09), following the RRab 
lists of \cite{molnar2018} and \cite{plachy2019}. The method of 
analysis was the same as for the Kepler data, but without the need 
for stitching and annual variation testing. The lack of successive 
campaign data on the same target, of course, requires higher level 
of caution, because we have no trustable information if a small 
amplitude variation on the time scale of the duration of the campaign 
is due to some seasonal effect or is a real variation in the star's 
amplitude (even though co-trending eliminates large part of the seasonal 
variation). 

Figure~\ref{K2_diagnostics} shows an example of the diagnostic plots 
obtained in the course the analysis of the K2 data. Various data sources 
show different degrees of systematics. By comparing the results from 
these sources not only indicate the effectiveness of the systematics 
filtering, but also increases the reliability of the final classification. 

%
%
\begin{table}[h]
\begin{flushleft}
\caption{Blazhko star statistics for K2 fields}
\label{k2_bl}
\scalebox{1.0}{
\begin{tabular}{rrccccc}
\hline
$C\#\#$   &   $M$ & $<eA/A>$ & $q0$ & $q2$  & $q1$ & $q12$ \\
\hline\hline
 1  &  14 & 0.00081 & 0.214 & 0.429 & 0.357 & 0.786 \\
 2  &  57 & 0.00055 & 0.035 & 0.579 & 0.386 & 0.965 \\
 3  & 124 & 0.00048 & 0.000 & 0.387 & 0.613 & 1.000 \\
 4  &  38 & 0.00093 & 0.211 & 0.368 & 0.421 & 0.789 \\
 5  &  72 & 0.00102 & 0.264 & 0.292 & 0.444 & 0.736 \\
 6  & 123 & 0.00043 & 0.081 & 0.350 & 0.569 & 0.919 \\
 7  & 246 & 0.00114 & 0.183 & 0.451 & 0.366 & 0.817 \\
 8  &  45 & 0.00042 & 0.089 & 0.289 & 0.622 & 0.911 \\
10  & 117 & 0.00053 & 0.214 & 0.274 & 0.513 & 0.786 \\
11  & 122 & 0.00034 & 0.090 & 0.393 & 0.516 & 0.910 \\
12  &  72 & 0.00046 & 0.097 & 0.431 & 0.472 & 0.903 \\
13  &  31 & 0.00084 & 0.129 & 0.290 & 0.581 & 0.871 \\
\hline
HQ :&  660& 0.00045 & 0.089 & 0.376 & 0.535 & 0.911 \\
LQ :&  401& 0.00106 & 0.197 & 0.401 & 0.401 & 0.803 \\
All:& 1061& 0.00068 & 0.130 & 0.385 & 0.484 & 0.870 \\
\hline
\end{tabular}}
\vskip 4 pt
{\bf Notes:}{\small\ 
$C\#\#$  = Campaign number; 
$M$      = Number of targets; 
$<eA/A>$ = Average of the relative amplitude errors (Eq.~\ref{eA/A});
$q0$, $q2$, $q1$, $q12$, respectively, are the fractions
of the class 0, 2, 1 and $1+2$ objects relative to $M$.  
The last three lines summarize the results for the 
high-quality ($C02$, 03, 06, 08, 10, 11, 12) and for the 
low-quality  ($C01$, 04, 05, 07, 13) fields. The relative 
amplitude errors in these lines are the averages of the 
campaign values, weighted by $M$.}
\end{flushleft}
\end{table}

Summary of the field-by-field observed occurrence rates for the various 
classes is given in Table~\ref{k2_bl}. As a data quality parameter, 
in column 3 we also give the averages of the relative amplitude 
errors, defined (somewhat arbitrarily) as follows 
%
%
\begin{eqnarray}
\label{eA/A}
eA/A = {\sigma \over A_t} \sqrt{2 \over N} \hspace{2mm}.  
\end{eqnarray}
Here $A_t$ denotes the total (peak-to-peak) fundamental mode amplitude, 
$\sigma$ is the standard deviation of the residuals after subtracting 
the full triplet model (see Sect.~\ref{sect:kepler}), and $N$ is the 
number of data points of the time series. When $A_t$ is substituted 
by one of the low-order Fourier amplitudes 
($A_k$ in $A_k\sin (k\omega+\varphi_k)$), Eq.~\ref{eA/A} gives the 
standard deviation ($1\sigma$ error) relative to that amplitude. 
The true relative error for $A_t$ is larger than $eA/A$, but this is 
unimportant in the present context.  

The inclusion of the last column -- the rate of all potential 
Blazhko stars -- is aimed for the indication of the compatibility 
of the present survey and our earlier, more limited surveys 
\citep{kovacs2018, kovacs2021}, where most of the class 2 objects 
were regarded as potential class 1 objects. 
 
We see that there is a broad positive correlation between $eA/A$ and 
$q0$ (the fraction of pure noise detections), whereas the correlation 
becomes negative for $q1$ (the fraction of the Blazhko detections). 
This is the expected effect of noise on the outcome of the time 
series analysis. The class 2 objects show no correlation. This 
indicates a mixture of objects with and without additional signals. 

The distribution of $eA/A$ shows an approximate bimodality: 
$eA/A > 0.0008$ for the 5 low-quality (LQ) fields whereas 
$eA/A < 0.0006$ for the 7 high-quality (HQ) fields. Using this 
bimodality, for the LQ fields we get $q1=0.401$, whereas 
for the HQ set we get $q1=0.535$. This seems to be a 
significant\footnote{By using Eq.~(1) of Alcock et al. (2003) 
for the sampling error -- assuming Poisson distribution -- we 
find that the difference between the two rates is significant 
above the 4 sigma level.} 
difference, underscoring the effect of observational noise on the 
derived occurrence rates.

%
%
%
\section{Occurrence Rates}
\label{sect:occur}
Considering only the two most important effects (noise and timebase) 
on the detection likelihood, we found it useful to compare the Kepler 
and K2 results with the unique analysis of \cite{prudil2017} and 
\cite{skarka2020} of the Galactic Bulge RR~Lyrae stars from the OGLE 
project. Because of the lack of strong evidence for the opposite, 
here we assume that the underlying distributions of the Blazhko stars 
are the same in the modulation frequency -- modulation amplitude space, 
independently of the population under investigation 
\citep[e.g., see][on the lack of metallicity effect in the frequency 
of Blazhko phenomenon]{smolec2005}. 

The accompanying supplementary material by \cite{skarka2020} allows 
us to compare the modulation properties on a common parameter space, 
involving the modulation frequency, $f_m$ and the side-lobe amplitudes,  
$A_{\pm}$ relative to the full pulsation amplitudes, 
$A_t$.\footnote{For the Galactic Bulge data, 
$A_t=(AmeanMIN+AmeanMAX)/2$ from Table~1 of \cite{skarka2020}.} 
To characterize the strength of the modulation and minimize the effect 
of blending, we use relative modulation amplitudes 
$A_s/A_t$, where $A_s=(A_{-}+A_{+})/2$. 

In the following subsections (Sects.~\ref{sect:mod_amp}, 
\ref{sect:mod_per}, \ref{sect:fmod-amod} and \ref{sect:inj}) we briefly 
review the observational support for the dependence of the observed 
occurrence rates on the data quality and time span. Finally, 
Sect~\ref{sect:u_rate} facilitates these observations in the estimation 
of the underlying occurrence rate for the K2 set. 

%
%
\subsection{The low modulation amplitudes}
\label{sect:mod_amp}
\begin{figure}[h]
\centering
\includegraphics[width=0.40\textwidth]{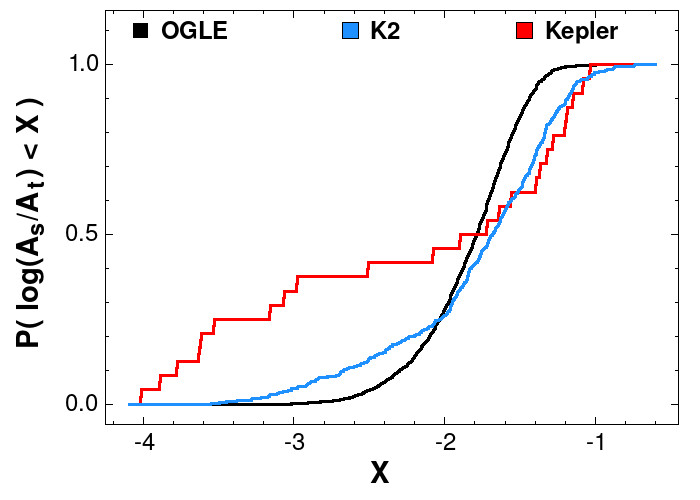}
\caption{Cumulative distribution function of the logarithm 
         of the relative side-lobe amplitudes from different 
	 data sources as given in the header of the figure. 
	 While OGLE samples the Galactic Bulge, Kepler and K2 
	 cover parts of the field.} 
\label{cdf_as}
\end{figure}

First we embark on the distribution of the modulation amplitudes. 
The cumulative distribution functions (CDFs) for the three datasets 
are shown in Fig.~\ref{cdf_as}. The Kepler set contains $24$ modulated 
stars. The high fraction of low modulation amplitudes ($33$\%, with 
$A_s/A_t < 10^{-3}$) is quite remarkable, even though the full 
RRab population is small, implying low/moderate statistical weight 
of the observed high fraction of low-amplitude modulations. 
In comparison, while the K2 sample contains $\sim 23$ stars ($4.5$\%), 
the OGLE sample has only $\sim 4$ ($0.12$\%). All these are in 
agreement with the expected trend of decreasing number of detections 
with the increase of the overall noise level of the data. 

In a separate observation we draw attention to the topological 
change in the CDF for K2 at $A_s/A_t=10^{-2}$\, $(X=-2)$, with 
an approximately linear decline toward smaller amplitudes. This 
behavior is quite different from what is observed for the OGLE 
sample. The change in the CDF of the K2 sample may indicate  
the existence of a subclass within the Blazhko stars below a 
critical modulation amplitude.

%
%
%
%

%
%
\subsection{The long modulation periods}
\label{sect:mod_per}
Turning to the modulation periods, $P_{BL}$, it is clear that 
aiming for long modulation periods with K2 is rather difficult, 
and, in fact, above $P_{BL}\sim 100$~d, we can only acknowledge 
the presence of the long-term variation, but even this can be 
done with a reasonable high confidence only if the modulation 
amplitude is large enough and the Blazhko phase is suitable 
at the time of the observation. Therefore, it is important to 
examine the fraction of long-period Blazhko stars in surveys 
with long baselines. With this information, we can guess the 
expected number of the long-period Blazhko stars lost in the 
K2 data, due to the short time spans of the various campaigns.    

%
%
\begin{table}[h]
\begin{flushleft}
\caption{Fraction of long Blazhko periods}
\label{long_bl}
\scalebox{1.0}{
\begin{tabular}{rrcllc}
\hline
$N_{tot}$ &   $N_L$ &    $R_L$ &   {\em Pop.} & {\em Data} &  {\em Source} \\
\hline\hline
 3141 &  871 & 0.384 & Bulge  & OGLE        &  [1] \\
  641 &  237 & 0.587 & Field  & SuperWASP   &  [2] \\
  310 &   98 & 0.462 & Field  & literature  &  [3] \\
  731 &  258 & 0.545 & LMC    & MACHO       &  [4] \\
   82 &   31 & 0.608 & M3     & literature  &  [5] \\
  514 &   36 & 0.075 & Field  & K2          &  [6] \\
   24 &    4 & 0.200 & Field  & Kepler      &  [6] \\
\hline
\end{tabular}}
\vskip 4 pt
{\bf Notes:}{\small\ 
$N_{tot}$ = total number of Blazhko stars; 
$N_L$ = Number of stars with $P_{BL}>100$~d; $R_L=N_L/(N-N_L)$; 
[1]=\cite{skarka2020}, 
[2]=\cite{greer2017}, 
[3]=the BlaSGalF site by Marek Skarka: \url{https://www.physics.muni.cz/~blasgalf/}, 
[4]=\cite{alcock2003},
[5]=\cite{jurcsik2019}, 
[6]=this paper.
See text for additional notes on the items in this table.  
}  
\end{flushleft}
\end{table}

Unlike for the modulation amplitudes, the available data 
in the literature on the modulation periods are fairly abundant. 
Including the Galactic Bulge data from OGLE, we found five 
extended sources covering various populations in the Galaxy 
and in the Large Magellanic Cloud (LMC). Table~\ref{long_bl} 
summarizes the relevant data, and most importantly $R_L$, 
the fraction of the long-periodic Blazhko stars to those 
with short modulation periods (i.e., with $P_{BL}<100$~d). 

The ``Data'' column typifies the main data acquisition method, 
with ``literature'' implying various ground-based telescopes, 
both individual, target-oriented projects and surveys, such 
as ASAS \citep{pojmanski2001}. The data on LMC and those in 
reference [3], with some extension were used also by \cite{skarka2016} 
to investigate the modulation period distribution of $1547$ 
RRab stars. These data yield $R_L=0.471$. The blind averaging  
of column 3 in the upper 5 rows yields $R_L=0.517$, whereas 
the averaging with $N_{tot}$ weighting results $R_L=0.443$, 
obviously dominated by the OGLE data. Omitting the likely 
specific rate for M3, simple averaging of the remaining four 
sources yields $R_L=0.495$. From these data we think that 
using a correction factor of $R_L=0.45-0.50$ is quite appropriate 
for the field RRab stars. With this factor we can estimate the 
long modulation period fraction in a Blazhko population, if only 
the short modulation period population is available due to 
observational constraints. 

%
%
\subsection{The $f_m$ -- $A_m$ map}
\label{sect:fmod-amod}
In addition to the marginalized distributions, it is useful to 
look at the data directly on the modulation 
frequency ($f_m$) -- modulation amplitude ($A_m$) plot. Although 
the real modulation amplitude could be several to many factors 
higher than the side-lobe amplitude at the fundamental frequency, 
both are useful for the characterization of the modulation strength. 
Therefore, we continue using $A_s/A_t$ also in the maps below. 

\begin{figure}[h]
\centering
\includegraphics[width=0.40\textwidth]{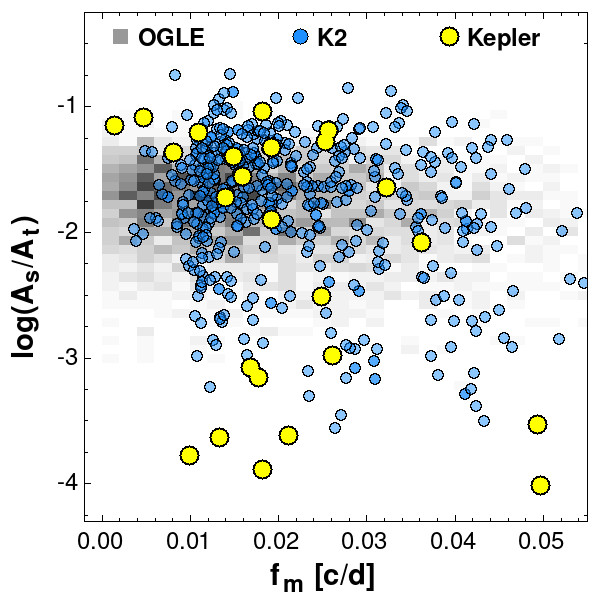}
\caption{Modulation frequency vs relative side-lobe amplitude 
         for the Galactic Bulge RRab inventory of the OGLE 
	 project \citep[3141 stars from][plotted on a 
	 gray-scale object number density map]{skarka2020}, 
	 the $514$ stars from $12$ K2 fields and the $24$ 
	 stars from Kepler's original field as detected in 
	 this paper. For comparison, there are only 4 stars 
	 in the OGLE sample with $A_s/A_t < 0.001$.} 
\label{fm_as_1}
\end{figure}

The detected Blazhko stars in the Kepler and in the twelve K2 fields 
are shown in Fig.~\ref{fm_as_1}. In comparison with the Galactic 
Bulge data in the background, one can observe two important 
differences. First, as already discussed earlier, Kepler and K2 
data (the latter naturally) miss many long-period Blazhko stars. 
Second, not entirely unexpectedly, the Kepler and K2 data reach 
much deeper toward low modulation amplitudes. Another relevant 
result from the OGLE data, is that although the survey is shallower 
in $A_m$ than Kepler, there does not seem to be a dependence on 
the modulation period. Therefore, the relative number of the 
detected long modulation periods ($R_L$ in Table~\ref{long_bl}) 
does not seem to suffer from observational bias. Furthermore, 
the Kepler and K2 amplitudes surpass the OGLE amplitude in both 
directions. We do not have an answer why do we have that also in 
the high-amplitude regime. The expected high level of deficiency 
of Blazhko stars with $f_m < 0.01$~c/d from K2 is clearly exhibited.  

%
%
%
%

%
%
\subsection{Detection limits from signal injection}
\label{sect:inj}
It is a basic question if the distribution of the amplitude modulation 
ends abruptly at low amplitudes or gradually shrinks to zero as the 
precision of the data allows detecting lower modulation amplitudes. 
In the former case, if the detection limit for a given dataset is known, 
one would see a clear break between the low-end of the distribution of 
$A_m$ and the detection limit. In a very preliminary way we looked at 
this question by performing injected signal tests in the Kepler and 
K2 data. 

Starting with the raw input fluxes \{$X_i$\}, we generated injected fluxes 
\{$Z_i$\} by modifying \{$X_i$\} according to the following formulae  
%
%
\begin{eqnarray}
\label{inj}
Y_i & = & 1 + A_{mod}\sin(\,2\pi\, f_{mod}\,(t_i-t_0)\,) \hskip 2mm , \\
Z_i & = & \overline X + (\,X_i-\overline X\,)\, Y_i \hskip 2mm ,
\end{eqnarray}
where \{$Y_i$\} is the modulation signal at the $i$-th data item, $A_{mod}$ 
is the total modulation amplitude (different from the side-lobe modulation 
amplitude $A_s/A_t$), $f_{mod}$ is the modulation frequency, and  
$\overline X$ is the robust average of \{$X_i$\}. Importantly, these 
modified fluxes go through the same steps (including systematics filtering) 
as the other, non-modified input time series (see Sect.~\ref{sect:method}). 
For any given injected time series, at fixed $f_{mod}$ and $t_0$, we looked 
for a proper value of $A_{mod}$ to reach the detection limit (the minimum 
of $A_{mod}$ at which the detection becomes secure, based on the diagnostics 
described in Sect.~\ref{sect:method}). With this $A_{mod}$, the analysis 
yields also the corresponding $A_s/A_t$, that can be compared with the 
observed values derived for the Blazhko stars. 

We selected non-Blazhko stars (class 0 and 2 stars) both from the Kepler 
set ($12$ stars) and from field $C08$ of K2 ($17$ stars). The Kepler set 
was tested at two modulation frequencies to investigate the detection 
sensitivity against long and short modulation periods (the former is to 
aid our understanding of the apparent miss of more long-period Blazhko 
stars in the Kepler set).

%
%
\begin{figure}[h]
\centering
\includegraphics[width=0.40\textwidth]{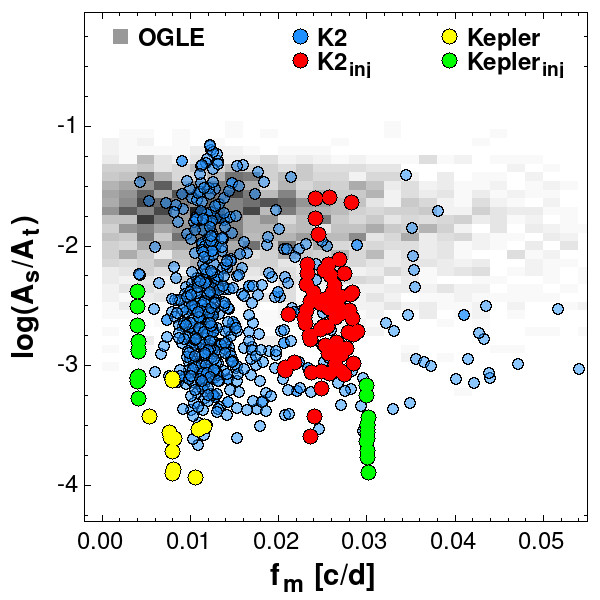}
\caption{Class 0 and 2 (i.e., non-Blazhko) stars from Kepler 
         and K2 on the modulation frequency -- relative side-lobe 
	 amplitude diagram (see footnote \ref{fm_am_class02}). 
	 For comparison, the Galactic Bulge Blazhko map is re-plotted. 
	 Also shown are the injection tests to assess detection efficacy. 
	 We used the non-Blazhko stars with the following injected 
	 modulation frequencies: 
	 $f_m=0.004$ and $0.03$~c/d for Kepler and 
	 $f_m=0.025$ for C08 of K2. The K2 injected result is plotted 
	 for all available data sources (i.e., there are multiple points 
	 for the same object).} 
\label{fm_as_2}
\end{figure}

Figure~\ref{fm_as_2} delivers the following, partially expected results 
for the Kepler data.\footnote{Because of the absence of modulation signals 
surpassing the detection criteria, for class 0 and 2 objects, the plotted 
quantities are simply those associated with the highest side peaks near 
the fundamental frequency.\label{fm_am_class02}} For the short-period  
injected signals the detection limits are close to the range of the 
low-amplitude Blazhko stars and also to the formal modulation amplitudes 
of the non-Blazhko stars (shown as yellow circles in Fig.~\ref{fm_as_2} 
-- see also footnote \ref{fm_am_class02}). This implies a continuous spectrum 
of modulation amplitudes down to the detection limit (i.e., no sign of 
abrupt stop of the Blazhko phenomenon below a critical modulation 
amplitude). For the modulation period approximately covering $2.5$ 
quarters, naturally, we have a higher detection limit, still, low-enough 
to reach Blazhko stars (even though this is a low-populated region according 
to the OGLE Bulge data). Although this may imply that the available  
small sample of RR~Lyrae stars in the Kepler field indeed lacks enough 
long-period Blazhko stars, it is important to recall that several stars are 
affected by annual variations (see Sects.~\ref{sect:method}, \ref{sect:kepler}), 
that may blur the long-period modulation. In any case, the hidden long-period  
Blazhko stars in the Kepler sample must have considerably lower modulation 
amplitudes than their counterparts in the Galactic Bulge. 

The K2 data obviously indicate a higher detection limit, but again, in 
agreement with the low-amplitude tail of the detected modulations. 
For the first sight, the size of the scatter in the detected modulation 
frequencies seems to be unexpected. However, considering the overall length 
of a campaign ($\sim 80$ days), the line widths of the peak profiles 
of the frequency spectra are in the range of $\sim 0.01$~c/d. Therefore, 
considering that these signals are at the detection limits, the observed 
scatter is not surprising.  

%
%
\subsection{Underlying occurrence rates}
\label{sect:u_rate}
If we forget about the sensitivity of the detection on the noise 
level (which is obviously incorrect), we can get a lower limit for  
the true number of Blazhko stars by simply considering the ratio 
between the short and long Blazhko periods as derived in 
Sect.~\ref{sect:mod_per} from samples, apparently complete in $P_{BL}$. 
In the K2 fields we found $514$ Blazhko stars. From these $36$ have  
$P_{BL}>100$~d. With a total number of $1061$ RRab stars in the sample 
and using a long-period boosting factor $0.45$ (see Sect.~\ref{sect:mod_per}), 
we arrive to a rate of $(514-36)\times 1.45/1061 = 0.653$ (or to 
$0.676$ if we use -- a still likely -- factor of $1.50$).  

For a better guess, we need to consider the fact that certain 
modulated stars might remain hidden even if $P_{BL}<100$~d, simply 
because of the high noise level in cases when the modulation 
amplitudes are low. Pursuing this deeper assessment of the 
underlying rate, we basically follow the method described in our 
former papers \citep[][]{kovacs2018, kovacs2021, kovacs2022}. 
In principle, this is a simple Monte Carlo (MC) simulation, by using 
the available ``best guesses'' on the distributions of the modulation 
periods and modulation amplitudes. Here we rely on the $A_s/A_t$ 
distribution derived in this work from the analysis of $12$ fields 
from K2 and the distribution for $P_{BL}$ as given by the OGLE data 
for the Galactic Bulge \citep[][]{skarka2020}. 

With the corresponding CDFs we can assign modulation frequencies 
to each member of the sample and inject modulation signals in the 
observed fluxes (in the same way as it was described in 
Sect.~\ref{sect:inj}, except that if the known Blazhko stars are 
included in the simulation, we would need to prewhiten their fluxes 
from the Blazhko component). This is obviously a formidable 
task, considering the large number of stars and the large number of 
randomly selected $(f_m,A_m)$ pairs needed for statistical stability. 
Instead, we use a simplified method (similar to those in our earlier 
works), enabling a quite close assessment of the detectability 
of a sinusoidal signal in the presence of noise.  

\begin{figure}[h]
\centering
\includegraphics[width=0.24\textwidth]{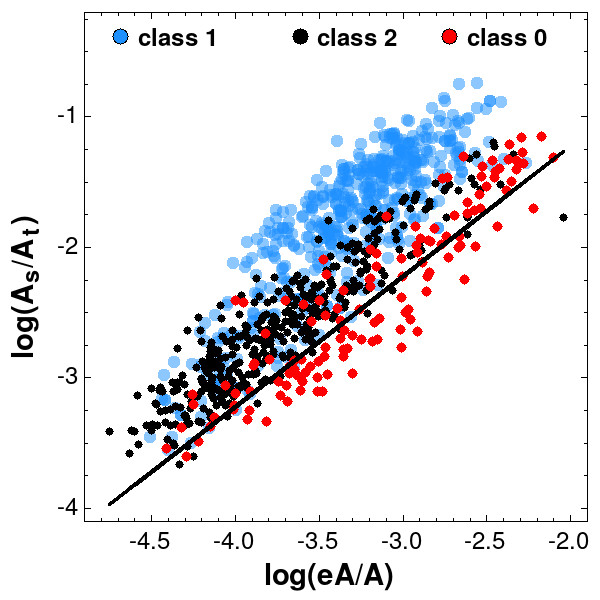}
\includegraphics[width=0.24\textwidth]{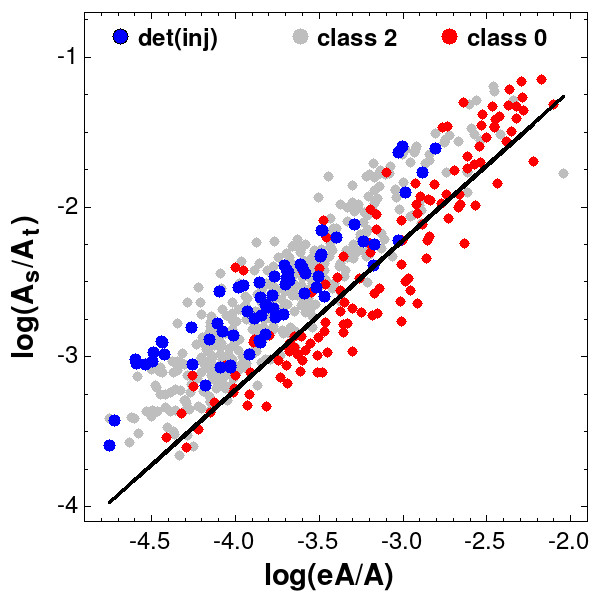}
\caption{{\em Left:} Relative amplitude error (see Fig.~\ref{eA/A})  
         vs relative side-lobe amplitudes for the $1061$ RRab 
	 stars from $12$ K2 fields. The straight line corresponds 
	 to $A_s/A_t=6\times eA/A$ and we call it the $6\sigma$ 
	 detection limit in this paper. 
	 {\em Right:} Injected signal test for the $17$ class 0 
	 and 2 stars in C08 of K2 (blue points). As in Fig.~\ref{fm_as_2}, 
	 we plot the results for all data sources used in the 
	 signal injection test.} 
\label{det_lim}
\end{figure}

To estimate the detection limit from the noise of the signal, we 
relate the relative amplitude error $eA/A$ (see Eq.~\ref{eA/A} in 
Sect.~\ref{sect:k2}) to the relative side-lobe amplitudes, $A_s/A_t$. 
We recall that $eA/A$ is given for all stars, due to the frequency 
analysis performed during the survey. We plot the detected side-lobe 
amplitudes as a function of the amplitude error in the left panel of 
Fig.~\ref{det_lim}. Although the correspondence is not entirely 
flawless, the $6\sigma$ criterion seems to be a reasonable detection 
limit: basically all class 1 and 2 objects fall above this detection 
limit, whereas a considerable number of the class 0 objects fall 
below.\footnote{The large jamming of class 0 object above the detection 
limit at high noise level is due to $C07$, the noisiest field. By leaving 
out this field, the separation of the class 0 objects becomes far 
more favorable.} 
The intermingling of the class 1 and 2 objects is also expected, 
because the latter do have Fourier signals, but their origin is 
unclear from the available data. 

%
%

It is also useful to check the performance of the above detection 
criterion for injected signals, where the analysis is performed 
independently of this criterion, relying entirely on the diagnostics 
described in Sect.~\ref{sect:method}. Using the same test result 
for the K2 field C08, as in Sect.~\ref{sect:inj}, we get the plot 
shown in the right panel of Fig.~\ref{det_lim}. Again, the $6\sigma$ 
condition looks valid, even though it was not used in tuning the 
injected amplitudes to the detectability. Furthermore, as partially 
expected, the detections cover the `gray area' of class 2 objects 
with Fourier detection flags but in the low/mid-SNR regime. 
 
Briefly, the search for the underlying occurrence rate yielding 
the best matching predicted rate to the one observed, is built 
on the following core loop. 

First, we randomly assign modulation status to $N_1$ members of 
the whole population of $N$ objects. Those with single-mode 
status ($N-N_1$ stars) are no longer dealt with in this particular 
loop. Next, based on the CDFs chosen for the generation of the 
modulation parameters, $(f_m,A_m)$ pairs are randomly assigned 
to the $N_1$ stars. Then, from the noise-detection relation 
described above, the detection status can be determined due 
to the knowledge of the start-by-star value of the observed 
relative noise level $eA/A$. Next, the detections with short 
modulation periods (i.e., those with assigned $f_m$ greater 
than some fixed lower limit $f_m^{min}$) are counted, and their 
number is registered as detections in the particular realization. 

The whole loop described above is repeated for an ample amount 
of times (in our case $500$ times). The derived number of 
detections in the $ir$-th realization, $N_{det}(ir)$ is used 
together with the values from the other realizations to compute 
the average and standard deviation of the predicted observable  
rate of short-period modulated stars for the assumed underlying 
rate of the full population of modulated stars (i.e., including 
those with $f_m<f_m^{min}$). The underlying rate is tuned until 
the observed and predicted rates match.   
   
Following a conservative approach, first we use the $P_{BL}$ 
distribution of the Bulge sample (although it yields lower rate 
boost than expected for the field stars) and native (i.e., derived 
from the K2 sample) modulation amplitude distribution (although 
the CDF derived from the Kepler data suggest a considerable 
higher boost).  

%
%
\begin{table}[h]
\begin{flushleft}
\caption{Underlying occurrence rates for K2}
\label{u_rates}
\scalebox{1.0}{
\begin{tabular}{lrcccccc}
\hline
{\em SET} & $N$ & $N_{BL}$ & $N_L$ & $Q^s_{obs}$ & $Q_{obs}$ & $Q_{und}$ \\
\hline\hline
LQ    &  401 &  161  &  15 & 0.364 & 0.402 & $0.590\pm0.019$ \\
HQ    &  660 &  353  &  21 & 0.503 & 0.535 & $0.780\pm0.017$ \\
All   & 1061 &  514  &  36 & 0.451 & 0.484 & $0.720\pm0.012$ \\
\hline
\end{tabular}}
\vskip 4 pt
{\bf Notes:}{\small\  
LQ, HQ, are the low- and high-quality subsets of K2 (see Table~\ref{k2_bl}); 
$N$, $N_{BL}$, $N_L$ are, respectively, the total number of stars, 
the number of Blazhko stars and those with $P_{BL}>100$~d; 
$Q^s_{obs}=(N_{BL}-N_L)/N$, $Q_{obs}=N_{BL}/N$ and $Q_{und}$ is the 
underlying rates of all Blazhko stars. The errors are the 
standard deviations of the values from the Monte Carlo simulations. 
CDFs from K2 and the Galactic Bulge were used to generate $(f_m,A_m)$ 
for the MC. See Sect.~\ref{sect:u_rate} for details of the MC simulations.}   
\end{flushleft}
\end{table}

Table~\ref{u_rates} summarizes the result for the K2 data. The short 
modulation period rates $Q^s_{obs}$ were matched with the predicted 
values within $\pm 0.002$. The overall boost in the incidence rates 
is $47$\%, shared between the boost considering the lost of class 1 objects 
due to noise (a boost with a factor of $1.15$ -- from the MC simulations 
without the $f_m<f_m^{min}$ constraint) and missing long-period 
modulations (a boost with a factor of $1.38$ -- since we used the 
Galactic Bulge sample to constrain the ratio of the long modulations). 
Knowing these factors, we can perform a brief sanity check of the 
derived underlying rates. For the full K2 dataset the number of detections 
with short modulation periods is $514-36=478$. Considering the boost 
due to noise bias gives $478\times1.15=550$. And, finally, the long-period 
bias yields: $550\times1.38/1061=0.715$, almost as shown in the 3rd line of 
Table~\ref{u_rates}.   

The above simple estimation can be extended for considering more realistic 
boost factors for the loss of long modulation periods. Using the lower 
value of $1.45$ from the preferred range of the modulation period bias 
factor (Sect.~\ref{sect:mod_per}), and staying with the moderate boost 
above for debiasing the noise effect, for the full K2 set discussed above 
we get $550\times1.45/1061=0.752$. 

Although the Kepler set has rather moderate statistical weight, it is 
still interesting to test the effect of changing the CDF from K2 for 
the modulation amplitude to the one derived from the Kepler set 
(see Fig.~\ref{cdf_as}). From the MC simulation, we get a factor of 
$\sim 1.5$ for noise debiasing. The expected rate from this and the 
boost for the long modulation periods derived from the field stars 
yield $478\times1.50\times1.45/1061=0.980$. A more sensible (albeit 
ad hoc) noise boosting factor of $1.25$ yields a rate of $0.82$. 
We conclude that current data for the K2 RRab sample support a 
true/underlying Blazhko occurrence rate of $75$\% or higher.

%
%
%
\section{Conclusions}
\label{sect:conclude}
In this paper we aimed at: (i) the deep analysis of the RRab stars in 
the Kepler field that offers a unique dataset to investigate these 
variables at the $100$~ppm level; (ii) extend the analysis to twelve 
K2 fields to compensate for the issues arising from the small sample 
size, and, more importantly, from the high level of crowdedness of the 
Kepler field; (iii) Revisit the question of the modulation amplitude 
distribution and the true occurrence rate of the Blazhko phenomenon. 
Concerning the last point, it is important to remark that apparently 
none of the works dealing with the occurrence rate of the Blazhko 
phenomenon are concerned much about observational biases, whereas 
these affect profoundly the number of detections. 

In the present study, there are several important ingredients that 
differ from our earlier works on the statistics of the Blazhko 
phenomenon \citep{kovacs2018,kovacs2021,kovacs2022}. Here is the 
list of the new features of the analysis presented in this paper. 

\begin{itemize}
\item[$-$]
We rely on the {\em observed} fraction of long modulation periods 
(from other surveys) and do not attribute every remnant Fourier peak   
near the fundamental mode to the effect of long-term modulation (as 
we largely did in our earlier works). 
\item[$-$]
For the K2 analysis we use up to four independent data sources, thereby 
greatly increasing the reliability of the detection. 
\item[$-$]
We use codes of higher sophistication, enabling us to perform better 
filtering of instrumental systematics and use various complimentary 
methods to search for light curve modulation. 
\item[$-$]
We use a three-level modulation classification scheme, where cases 
of ``significant spectral residual only'' do not count, if they 
are not accompanied by other supporting diagnostics.  
\end{itemize}
 
The main conclusions of this paper can be summarized as follows. 

\begin{itemize}
\item[$-$]
Although the Kepler data on RR~Lyrae stars is at the highest standards  
both in length and quality, the high crowding and the accompanying 
instrumental effects make it difficult to aim at long-period Blazhko 
modulations at the ppt level or lower. 
\item[$-$]
In spite of the main obstacle above, we found $7$ new Blazhko stars with 
modulation periods less than $\sim 100$~d, and side-lobe amplitudes 
around $250$~ppm (for $6$ stars). The considerable fraction of $6/24$ 
of such a low-amplitude Blazhko stars is a warning for other, shallower 
surveys for missing a non-negligible fraction of the Blazhko population 
\item[$-$]
From the literature, covering various RR~Lyrae populations and 
instrumentation, we found that the relative number of Blazhko stars 
with long modulation periods (i.e., with $P_{BL}>100$~d) is very high. 
In the overall sense, there are half as many Blazhko stars in the 
long-period regime than in the short-period regime. This, again, a 
warning sign for the short-term surveys when counting the Blazhko 
population.  
\citep[e.g., works based on the non-continuous viewing sectors of the TESS satellite --][]{molnar2022}. 
\item[$-$]
From the analysis of $1061$ RRab stars in $12$ K2 fields yields an 
observed occurrence rate of $48$\%, which, after correction for 
biases by the finite observational timebase and noise, increases 
to $75$\%. This is our best (although still conservative) estimate 
for the underlying occurrence rate of Blazhko RRab stars in the 
Galactic field. 
\item[$-$]
Based on the analysis of the Kepler and K2 data, the modulation 
amplitudes do not seems to exhibit a minimum value; there are 
modulations close to the detection limits in both datasets, 
continuously extending the distribution to very low amplitudes. 
\end{itemize}

%
%
\begin{acknowledgements}
%
We appreciate Marek Skarka's quick response on the parameterization of the 
OGLE's Bulge RR~Lyrae sample. We thank Eric Petigura, Andrew Vanderburg 
and Rodrigo Luger for answering questions related to data accessibility.  
The quick and instructive report of the referee is acknowledged. 
This paper includes data collected by the Kepler mission and obtained 
from the MAST data archive at the Space Telescope Science Institute 
(STScI). Funding for the Kepler mission is provided by the NASA 
Science Mission Directorate. STScI is operated by the Association 
of Universities for Research in Astronomy, Inc., under NASA contract 
NAS 5-201326555.

%
%
%
%
%
\end{acknowledgements}

%
%

%

%
\end{document}